\documentclass[useAMS,usenatbib]{mn2e}
\usepackage{times}
\usepackage{graphicx}
\usepackage{amssymb}
\usepackage[fleqn]{amsmath}

\makeatletter

\newcommand{\yr}{{\,\rm yr}}

\newcommand{\pc}{\,\mathrm{pc}}
\newcommand{\kpc}{\,\mathrm{kpc}}

\newcommand{\Mbh}{M_{\bullet}}

\newcommand{\Mo}{M_{\odot}}

\newcommand{\Ms}{M_{\star}}

\newcommand{\tp}{t_{\omega}}
\newcommand{\Jp}{J_{\omega}}

\makeatother

\title[Resonant relaxation]{Resonant relaxation near a massive black hole: the dependence on eccentricity}
\author[M. A. G\"urkan and C. Hopman]{M. Atakan G\"urkan$^{1}$\thanks{E-mail:
ato@science.uva.nl (MAG); clovis@strw.leidenuniv.nl (CH)} and Clovis
Hopman$^{2}$\footnotemark[1]\\
$^{1}$Astronomical Institute ``Anton Pannekoek'',
University of Amsterdam, Kruislaan 403, 1098 SJ Amsterdam, The Netherlands\\
$^{2}$Leiden University, Leiden Observatory, P.O. Box 9513,
NL-2300 RA Leiden, The Netherlands}
\begin{document}
\bibliographystyle{mn2e.bst} 

\date{Accepted 200x XXX xx. Received 200x XXX xx; in original form 200x XXX xx}

\pagerange{\pageref{firstpage}--\pageref{lastpage}} \pubyear{200x}

\maketitle

\label{firstpage}

\begin{abstract}
The orbits of stars close to a massive black hole are nearly
Keplerian ellipses. Such orbits exert long term torques on each
other, which lead to an enhanced angular momentum relaxation known as
resonant relaxation.  Under certain conditions, this process can
modify the angular momentum distribution and affect the interaction
rates of the stars with the massive black hole more efficiently
than non-resonant relaxation.  The torque on an orbit exerted by
the cluster depends on the eccentricity of the orbit. In this paper,
we calculate this dependence and determine the resonant relaxation
timescale as a function of eccentricity. In particular, we show that
the component of the torque that changes the magnitude of the angular
momentum is linearly proportional to eccentricity, so resonant
relaxation is much more efficient on eccentric orbits than on
circular orbits.
\end{abstract}


\section{Introduction}\label{s:intro}

It is now commonly accepted that many galaxies contain massive black
holes (MBHs) at their centres \citep[e.g.,][]{Geb03, Mil06}, with
masses $10^6\Mo\!\gtrsim\!\Mbh\!\gtrsim\!10^9\Mo$. The proximity of
our own galactic centre, at a distance of $7.62\pm0.32\kpc$
\citep{Eis05}, allows for the astrometric study of individual stellar
orbits in the potential well of the MBH \citep[e.g.,][]{Sch02, Ghe03b,
Eis05}. These observations showed that the galactic MBH mass is
$\Mbh=(3.61\pm0.32)\times10^6\Mo$ \citep{Eis05}. The presence of
very dense stellar clusters near MBHs leads to a wide variety of
interesting phenomena. For a comprehensive review of stellar phenomena
in the Galactic centre, see \citet{Ale05}. Many of these phenomena
arise because the two-body relaxation time is shorter than the age of
the systems, so that the orbits of stars are significantly
redistributed during their lifetimes. For example, the relaxation
time in the Galactic centre is estimated to be several Gyr
\citep{Ale99a, Ale03b}.

A common assumption in stellar dynamics is that the mechanism through
which stars exchange angular momentum and energy is dominated
by uncorrelated two-body interactions \citep[e.g.,][]{Cha43}.
The orbits of the stars are largely determined by the potential
of the smoothed density and the deviations from this potential
caused by the individual stars lead to perturbations that evolve the
orbits \citep{Hen73}.  Systems with MBHs at their centres cannot (yet) be
studied by direct $N$-body integrations using a realistic number of
particles. Hence the methods used for studying the dynamics of these
systems rely on the above assumptions. Such studies include Monte
Carlo simulations \citep[e.g.,][]{Sha78, Fre01, Fre02, Fre03, Fre06}
and Fokker-Planck methods \citep[e.g.,][]{Bah76, Bah77, Mur91, Hop06b}.

Near a MBH, the potential is nearly Keplerian, which leads to closed
elliptic orbits. 
There are two main reasons for deviations from Keplerian orbits:
the contribution to the potential from the stars and general
relativistic effects.
However, for most
orbits where the potential is dominated by the MBH, the time scale for
precession is large, and the orbits remain nearly stationary with
respect to each other over many periods. Consequently, the assumption
that the interaction between the stars are independent is not valid,
as shown by \citet{Rau96}.  They argued that a better description is
given when the interactions are considered to be between different
{\it orbits}, rather than between different {\it point
particles}. Averaged over many periods, one can think of the mass of a
star being smoothly distributed over the orbit, with the linear
density in a small segment proportional to the time the star spends in
this segment. The orbits then form massive ``wires''. \citet{Rau96}
showed that the torques between the wires lead, under some
circumstances, to very efficient angular momentum relaxation.

Part of the interest in efficient angular momentum relaxation, is due
to interesting phenomena which occur when stars have very close
interactions with MBHs when they are on highly eccentric orbits. Such
phenomena include the tidal disruption of stars \citep[e.g.,][]{Fra76,
Lig77, Ree88}, and the emission of gravitational waves with detectable
frequencies \citep[e.g.,][]{Sig97a, Fre03, Hop05, Hop06}. Since
resonant relaxation increases the rate of angular momentum scattering,
stars reach highly eccentric orbits more rapidly. \citet{Rau98}
studied the consequences for the tidal disruption rate, while
\citet{Hop06} considered the enhancement of the rate at which
gravitational wave sources spiral in. Resonant relaxation appears to
be more important for the latter case, since it occurs closer to the
MBH, where deviations from a $1/r$ potential are smaller.

Resonant relaxation affects both the magnitude and the direction of
angular momenta. For processes that require orbits of high
eccentricity, only the part which affects the overall magnitude of
the angular momentum is of interest. \citet{Rau96} estimate the
torques of a stellar cluster on a test star, and deduce the relaxation
rate from this. They do not consider the dependence of this process on the
eccentricity. This is potentially important, as can be
seen from the following example.

Consider a star in a circular orbit in the $xy$-plane, and a mass at a general
point $\vec{p}=(x,y,z)$. In this case, the angular momentum of the
star is in the $z$ direction, $\vec{J}=J\hat{z}$. In the approximation
that the orbit does not change, the torque
$\vec{\tau}$ that the mass at $\vec{p}$ exerts on the star would have no $z$
component since the contributions from one half of the orbit will cancel
the contributions from the other half, $\vec{\tau}=(\tau_x, \tau_y, 0)$.
Since $\vec{\tau}{\mathbf \cdot}\vec{J}=0$, the mass at $\vec{p}$ can
rotate the orbit, but never affect the magnitude of the angular
momentum, and since $\vec{p}$ was a general point, this is true for
all other points as well. We can therefore conclude that resonant
relaxation can never modify the eccentricity of an $e=0$ orbit.

For any orbit with $e>0$, resonant relaxation will change the
eccentricity. In this paper, we calculate how the efficiency of
resonant relaxation depends on the eccentricity of the orbit of a given
star.

\section{Resonant relaxation}\label{s:rr}
In a Keplerian orbit there is a $1\mathbin{:}1$ resonance between the angular and radial frequency and hence the orientation of the orbit is fixed
in space.
Because of the general relativistic effects, all orbits around a
MBH actually precess and are never exactly Keplerian;
in addition, the contribution to the potential from the star
cluster around the MBH will also lead to precession.
However, for nearly Keplerian systems,
the orientation of the orbit with respect to other orbits
can be assumed to be fixed in space over some time $\tp\gg P$, where

\begin{equation}\label{e:P}
P(a)={2\pi}\left(\frac{a^3}{G\Mbh}\right)^{1/2}
\end{equation}
is the period of the star.

Over a timescale $P\!\ll\! t \!\ll\! \tp$, 
the stars can be represented as massive wires \citep{Rau96}
with the mass smeared out over their orbits. These wires exert mutual
torques on each other. The magnitude of the torque exerted by a star
of mass $\Ms$ and semi-major axis $a$ on another star with equal
semi-major axis is estimated by
\footnote{For brevity, we use angular momenta, torque etc. per unit mass.}

\begin{equation}\label{e:torque1}
\tau_{1}\sim \frac{G\Ms}{a}.
\end{equation}

For a large number $N$ of stars in the region near\footnote{In Section
\ref{s:method} we show that the maximal distance at which stars still
have a large contribution to the torque is $2a$.}  $a$, the sum of the
torques will nearly average out to zero.  However, because of
statistical fluctuations, there will be an excess torque in an unknown
direction of order $\tau_{N}\sim\sqrt{N}\tau_{1}$. If the orbit of the
test star lies in the $xy$-plane, so that its angular momentum is in
the $\hat{z}$-direction, only the $z$-component of the torque can
affect the magnitude of the angular momentum (or the eccentricity) of
the star. The resulting angular momentum changes were called {\it
scalar} resonant relaxation by \citet{Rau96}. If the eccentricity of a
test stars is $e$, then the typical net torque in the $z$-direction
will be

\begin{equation}\label{e:torqueN}
\tau_{z}(a, e) = \beta_s(e)\sqrt{N}\frac{G\Ms}{a},
\end{equation}
where $\beta_s(e)$ is a dimensionless function of eccentricity. 

In \citet{Rau96} and later papers, the eccentricity dependence of $\beta_s$ 
was ignored. \citet{Rau96} performed $N$-body
simulations to find the eccentricity averaged value of $\beta_s(e)$,
and found that for an isotropic cluster with eccentricity distribution

\begin{equation}\label{e:iso}
N_{\rm iso}(e)=2e\,de,
\end{equation}
the typical value of $\beta_s$ is $\bar{\beta}_s = 0.53\pm0.06$
\citep[][table 4d]{Rau96}. However, from the discussion in the
introduction, it follows that $\beta_s(0)=0$.

After a time $\tp$, the orientation of the test star with respect to
the ambient cluster changes. This can happen either because its orbit
has precessed, or because the orbits of most of the stars around it
have precessed. Over this time, the angular momentum would change by

\begin{equation}\label{e:Jw}
\Delta\Jp=\dot{J}\tp= \beta_s(e)\sqrt{N}\frac{G\Ms}{a}\tp(a,e)\,.
\end{equation}

For $t\!>\!\tp$ the torques on a particular star-wire become random,
and the change in angular momentum grows as a random walk. The
resonant relaxation time $T_{\rm RR}$ is defined as the time is takes
for a star to have its angular momentum changed by an amount of the
circular angular momentum

\begin{equation}\label{e:Jc}
J_c(a) = \sqrt{G\Mbh a}\,.
\end{equation}
Since it takes $(J_c/\Delta\Jp)^2$ random steps to make this excursion
in angular momentum space, and each step takes a time $\tp$, the
resonant relaxation time is given by 

\begin{equation} \label{e:TRR}
\begin{split}
T_{\rm RR}(a,e)&= \left(\frac{J_{c}}{\Delta J_{\omega}}\right)^{2}\tp\\
               &= \left[\frac{1}{2\pi\beta_s(e)}\right]^2
	          \left(\frac{\Mbh}{\Ms}\right)^2
                  \frac{1}{N(a)}
                  \frac{P(a)^2}{\tp(a,e)}.
\end{split}
\end{equation}

In the following section we describe a method to find the torque on a
wire which we use to determine the eccentricity dependence of
$\beta_s$. We note that from equation (\ref{e:TRR})
is can be seen that since $\beta_s(e)\to0$ for $e\to0$, scalar resonant
relaxation becomes very ineffective for nearly circular orbits.

\section{The wire approximation for torque computation}\label{s:code}
\label{s:method}
For our computations, we use a simple model that describes the
Galactic centre. We assume $\Mbh=3.6\times10^6\Mo$ and $\Ms=\Mo$ for
the mass of the MBH and the mass of each star, respectively.  The
radius of influence of the MBH, where the mass in stars is equal to
$\Mbh$, is $r_h=2\pc$, and within this distance from the MBH
there is a cusp of stars,
\begin{equation}\label{e:cusp}
N_{\rm cusp}(\mathord{<}a)=N_h \left(\frac{a}{r_h}\right)^{3-\alpha},
\end{equation}
where $N_h=\Mbh/\Ms$ is the number of stars within the radius of
influence $r_h$, and $\alpha$ is the slope of the number density profile, for
which we adopt the value $\alpha=1.4$ \citep{Ale99a, Gen03a,
Ale05}. 

In order to compute the torque from a cluster of stars on a given test
star efficiently, we make use of the ``wire approximation'' suggested
by \citet{Rau96}. We consider a test star in an orbit of some given
initial eccentricity $e_{\rm t}$ and semi-major axis $a_{\rm t} =
0.01\,{\rm pc}$ in the $xy$-plane, with the orbit's peri-apse on the
positive $x$-axis.  It is surrounded by a cluster of field stars whose
eccentricities are drawn randomly from an isotropic distribution
$N_{\rm iso}(e)$ (see Eq. [\ref{e:iso}]) and semi-major axes from the
distribution given in equation (\ref{e:cusp}).

We truncate the semi-major axes of the field stars at
$5a_{\rm t} = 0.05\,{\rm pc}$, giving $N=10000$ stars.
The orbits for the field stars start in a configuration similar to
the test star and undergo a number of rotations: They
are first rotated around the $z$-axis by an angle $\phi$, drawn from
a uniform distribution in $[0, 2\pi)$; then rotated around their
latus rectum by an angle $\theta$, cosine of which is
drawn from a uniform distribution in $[-1,1]$; and finally rotated
around their major axis by an angle $\gamma$, drawn from a
uniform distribution in $[0, 2\pi)$. For a point starting from $(x,
y, 0)$ the final coordinates $(x', y', z')$ are given by 

\begin{equation}
\begin{split}
x' =& -y\cos\phi \sin\theta \sin\gamma \\
    & +y\sin\phi(2\cos^2\phi\cos\theta - \cos2\phi)\cos\gamma\\
    & +x\cos\phi(\cos2\phi\cos\theta + 2\sin^2\phi);
\end{split}
\end{equation}

\begin{equation}
\begin{split}
y' =& -y\sin\phi\sin\theta\sin\gamma \\
    & +y\cos\phi(2\sin^2\phi\cos\theta + \cos2\phi)\cos\gamma\\
    & +x\sin\phi(\cos2\phi\cos\theta - 2\cos^2\phi);\\
\end{split}
\end{equation}

\begin{equation}
\begin{split}
z' =& -y(\cos\theta\sin\gamma+2\cos\phi\sin\phi\sin\theta\cos\gamma)\\
    & -x\cos2\phi\sin\theta
\end{split}
\end{equation}

To calculate the torques, we represent the orbits by discrete points
that are equidistantly spaced in the mean anomaly of the orbit. We
start by using 64 points on each orbit to calculate the torque. We
estimate the error in our calculation by recomputing the torque with
the points which are in the middle (in mean anomaly) of the points
just used. If the relative difference between the two torques
calculated, $\delta_\tau = \left|(|\vec\tau_1| - |\vec\tau_2|) /
(|\vec\tau_1|+|\vec\tau_2|)\right|$, is larger than 0.01, we double
the number of points and repeat the computation. Once the desirable
tolerance is reached, we use $\vec\tau = (\vec\tau_1 + \vec\tau_2)/2$
for torque. To limit the time spent for computation, we use at most
65536 points per orbit. When we quadrupled this value during the test
runs, we obtained virtually identical results. Typically, a few
thousand points per field star were required.

\section{Results}\label{s:results}

\subsection{The torque}\label{ss:torque}
To determine the eccentricity dependence of torque on an orbit, we
made simulations with $e_{\rm t}$ = 0, 0.2, 0.4, 0.6, 0.8, 0.99 and 0.999.
For each value of $e_{\rm t}$, we carried out 80 simulations with 
$N=10000$ stars each and averaged over the results. 

\begin{figure}
\includegraphics[width=\hsize]{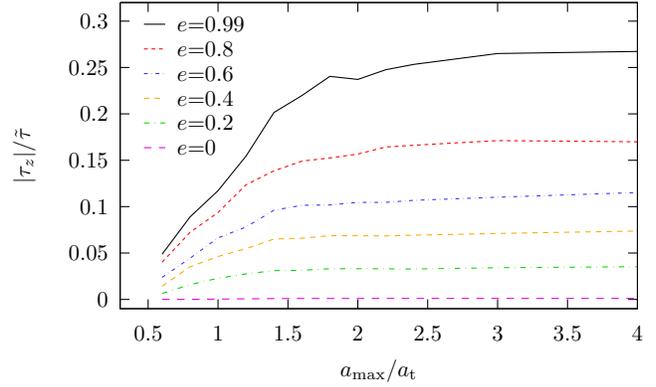}
\caption{\label{f:amax_dep}
The $z$ component of the torque computed from stars with semi-major
axes smaller than a given value, $a_{\rm max}$.  The flattening of
the curves implies that contribution from stars with semi-major axes
much larger than the test star's is negligible. Results for $e=0.999$
are very close to $e=0.99$ case and are not shown here.
}
\end{figure}

As a first result, we confirmed that the total torque is proportional
to $N^{1/2}$. However, stars very far from the test star will not
exert any discernible torque. We therefore first determine beyond
which point the contribution from stars become negligible.  In Fig.
\ref{f:amax_dep}, we plot the $z$ component of the torque on a star
of semi-major axis $a_{\rm t}$ as a function of $a_{\rm max}$, where
$a_{\rm max}$ is the cut-off for the semi-major axes of the field
stars. This figure shows that stars with semi-major axis larger than
the test star's apo-centre distance $r_{\rm apo}=a_{\rm t}(1+e_{\rm
t})$, contribute very little to the net torque on the test
star. Motivated by this, we normalize the torque by
\begin{equation}
\tilde{\tau} = \sqrt{N(\mathord{<}2a_{\rm t})} \frac{G \Ms}{a_{\rm t}}\,.
\end{equation}

\begin{figure}
\includegraphics[width=\hsize]{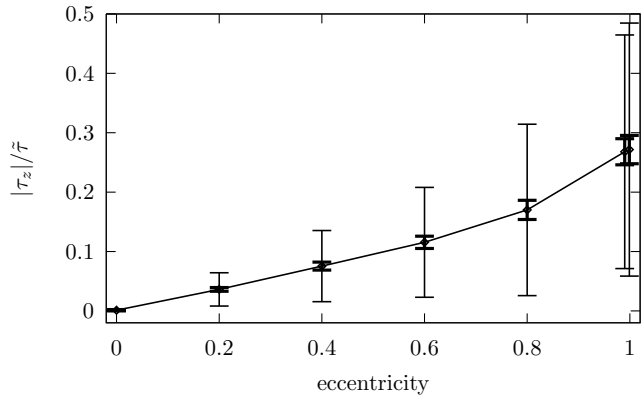}
\caption{\label{f:tau_z-vs-ecc} The $z$ component of torque (parallel
to the star's angular momentum) as a function of eccentricity. The
large error bars give an estimation of the root mean square variations
of the torques. These variations can be of order unity, so that the
torque of a star with given eccentricity can vary considerably
depending on the ambient stellar cluster; the torque for a given
configuration is expected to lie within the large error bars. The
small error bars estimate the uncertainty of the {\it average} torque
of a given eccentricity, based on 80 different configurations of the
host cluster. The average torque is well determined: if it were to be
computed again from 80 different cluster configurations, it is
expected to lie within the small error bars.}
\end{figure}

Fig. \ref{f:amax_dep} already shows a strong dependence of the $z$
component of the torque on eccentricity. In Fig.
\ref{f:tau_z-vs-ecc} we show this more explicitly by plotting the
$z$ component of the torque as a function of eccentricity. As
expected, the torque vanishes for $e\to0$, and has finite values for
$e>0$. We find that the result is consistent with a linear growth of
the torque as a function of $e$.  The best linear fit gives

\begin{equation}\label{e:beta_s}
\tau_z = \beta_s(e)\,\tilde{\tau}= 0.25\,e\,\tilde{\tau} \,.
\end{equation}
In Fig. \ref{f:tau_z-vs-ecc}, a cusp with $\alpha=1.4$ was
assumed. We have performed another set of calculations in which
$\alpha=2$, which also showed a linear eccentricity dependence of
$\tau_z$. It can thus be concluded that the result in equation
(\ref{e:beta_s}) is not strongly depended on the particular choice of
$\alpha$.

\begin{figure}
\includegraphics[width=\hsize]{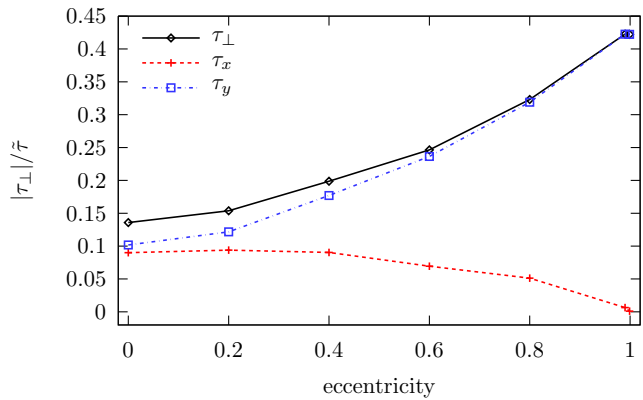}
\caption{\label{f:tau_perp-vs-ecc}
The components of the torque perpendicular to angular momentum, as
a function of eccentricity.  For clarity, we do not plot the error
bars in this figure, but they are comparable to the ones shown in
Fig. \ref{f:tau_z-vs-ecc}.
}
\end{figure}

The component of the torque perpendicular to the angular momentum,
$\tau_{\perp}\equiv\sqrt{\tau_x^2+\tau_y^2}$ changes its direction but
not magnitude. The resulting relaxation process is called {\it vector}
resonant relaxation by \citet{Rau96}. We plot this component as a
function of eccentricity in Fig.~\ref{f:tau_perp-vs-ecc}. Here, the
data are consistent with the torque being a quadratic function of
eccentricity:
\begin{equation}\label{e:beta_v}
\tau_\perp = \beta_v(e)\,\tilde{\tau}= 0.28\,(e^2+1/2)\,\tilde{\tau} \,.
\end{equation}
Note that the $x$ component of the torque vanishes for large $e$, and
$x$ and $y$ components become equal to each other for small $e$.

\subsection{The resonant relaxation time}\label{s:trr}
For the translation of the torques into a relaxation time, the
time-scale for the stars to change their orientation with respect to
the host cluster $\tp$ in Eq. (\ref{e:TRR}) needs to be
determined. Three relevant processes are (1) precession of the test
star due to general relativity; (2) precession of the test star due to
the extended distribution of the host cluster and (3) precession of
the host cluster itself due to its own extended distribution.

Torques are assumed to be coherent when the orbit has precessed less
than an angle $\omega$, and to make a random walk for angles
$>\!\omega$. The precise value of the coherence angle $\omega$, which
determines the steps of the random walk, is not clear. In particular,
the coherence angle may itself depend on eccentricity,
$\omega=\omega(e)$. Our wire method, which does not include the
evolution of the orbits, is not well suited to determine this
dependence, and we do not consider this possibility here. By rotating
a test wire in a cluster, we find that typical variations of the
torque occur over angles of $\sim\pi/2$. Motivated by this, we use
$\omega=\pi/2$, smaller than the value of $\omega=\pi$ assumed by
\citet{Rau96}. We note that a larger value for $\omega$ leads to more
effective resonant relaxation. We now discuss the three processes leading
to reorientation of the test star's orbit with respect to the cluster.

The general relativistic precession time is given by
\begin{equation}\label{e:tGR}
t_{\rm GR}(a,e)={\frac{4}{3}}\left(\frac{J}{J_{\rm LSO}}\right)^{2}P = 
  \frac{a}{12r_S}(1-e^2)P(a),
\end{equation}
 where

\begin{equation}
J_{\rm LSO}\equiv{\frac{4G\Mbh}{c}}\label{e:JLSO}
\end{equation}
is the angular momentum of the last stable orbits for eccentric orbits,
and $r_S = 2G\Mbh/c^2$ is the Schwarchild radius of the black hole.

Since the potential is not exclusively dominated by the MBH, but there
is a contribution of the stellar cluster as well, the orbit of the test star
precesses. The precession rate of a star in a cusp near a MBH was
derived by \citet{Iva05}. Here we briefly summarize the result for
$\alpha=3/2$. Let $\delta\omega(a,e)$ be the change in the angle of
the peri-centre during one orbit. The timescale for precession due to
extended mass distribution is then given by

\begin{equation}\label{e:tprec}
t_{M}(a,e)=\frac{\pi}{2\delta \omega(a,e)}P(a).
\end{equation}
 \citet{Iva05} showed that 

\begin{equation}
\delta \omega(a,e) = 4\left(\frac{a}{r_h}\right)^{3/2}{\sqrt{1-e^2}\over 2e}{d\over d e}F(e),
\end{equation}
where
\begin{equation}
F(e) = {2\over3}\sqrt{(1+e)}\left[4E\left(\sqrt{2e\over 1+e}\right) - (1-e)K\left(\sqrt{2e\over 1+e}\right)\right]\,,
\end{equation}
and $E$ and $K$ are complete elliptic integrals.
For this result, an $\alpha=3/2$ power-law was assumed, which
simplifies the equations. For more general expressions see
\citet{Iva05}.

Since general relativistic precession takes place in the opposite
direction, the rate at which the star's orbit precesses due to the combined
effects of general relativity and the extended potential is
\citep{Hop06}

\begin{equation}\label{e:tos}
\tp^{*}(a,e) = \left| {1\over t_{\rm GR}(a,e)} -  {1\over t_{M}(a,e)}\right|^{-1}.
\end{equation} 

For some combinations of $(a,e)$, the precession time $\tp^{*}(a,e)$
can become very large, implying that the star does not precess with
respect to inertial space. However, for the efficiency of resonant
relaxation, it is the orientation of the orbit {\it with respect to the
other stars} that matters. If most of the other stars do precess, the
torque on the test star will still fluctuate. We therefore define the
precession time of the stellar cluster as

\begin{equation}\label{e:tocl}
\tp^{\rm cl}(a) \equiv \tp^{*}(a,e=0.7).
\end{equation} 
The eccentricity $e=0.7$ is the median eccentricity for an isothermal
eccentricity DF; approximately half of the star precess more rapidly
than $\tp^{*}(a,e=0.7)$, and half of the star experience slower
precession.

The limiting time scale for resonant relaxation is then

\begin{equation}\label{e:tp}
\tp(a,e) = \min\left[\tp^{*}(a,e), \tp^{\rm cl}(a)\right].
\end{equation} 

Using equation (\ref{e:tp}) in equation (\ref{e:TRR}) gives the
resonant relaxation time as a function of $a$ and $e$. 

\subsection{Resonant relaxation for a simple model of a galactic nucleus}\label{s:GC}

We apply our results to a simple model which may describe a galactic
nucleus similar to our Galactic centre. For masses, we assume that
$\Mbh=3.6\times10^6\Mo$ and $\Ms=\Mo$. The radius of influence of the
MBH, where the mass in stars is equal to $\Mbh$, is $r_h=2\pc$, and
there is a cusp of stars, with $\alpha=3/2$ (see Eq. \ref{e:cusp}).

\begin{figure}
\includegraphics[angle=270,width=\hsize]{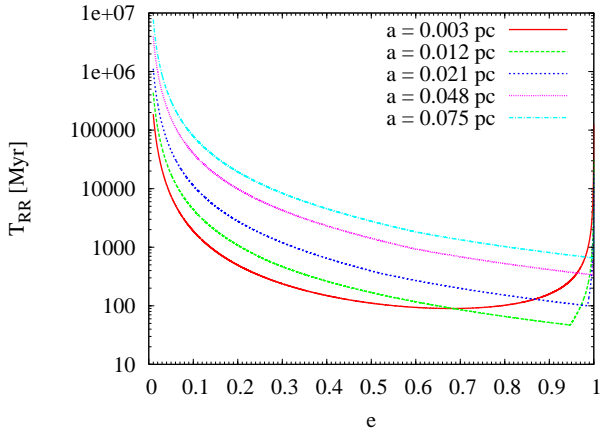}
\caption{\label{f:Te}Resonant relaxation time $T_{\rm RR}$ as a
function of eccentricity $e$, for several examples of semi-major axis,
for the parameters of the Galactic centre. For small $e$, $T_{\rm
RR}\to\infty$, because $\beta_s(e)\to0$. The resonant relaxation time
decreases with $e$, mainly because the torques $\tau_{1}\propto
e$. For large $e$ and small $a$, $T_{\rm RR}$ increases again with
$e$, because the precession time becomes very short due to general
relativistic effects. This effect is not of importance for very large
semi-major axes.}
\end{figure}

In Fig. \ref{f:Te} we show $T_{\rm RR}(a,e)$ as a function of $e$
for several choices of $a$. For small eccentricities, $T_{\rm RR}$
becomes very large, and non-resonant relaxation is much more effective
in changing the angular momenta than resonant relaxation. The resonant
relaxation time then decreases with $e$, but for small semi-major axes
it increases again near $e=1$, because general relativity causes
rapid precession of the orbit (Eq. \ref{e:tGR}).

\begin{figure}
\includegraphics[angle=270,width=\hsize]{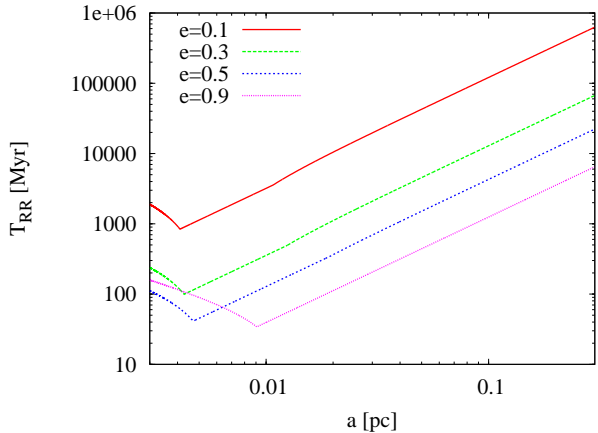}
\caption{\label{f:Ta}The resonant relaxation time $T_{\rm RR}$ as a
function of semi-major axis $a$, for several choices of
eccentricity. Far away from the MBH, the $T_{\rm RR}$ increases with
distance, but it reaches a minimum near $0.01\pc$, where general
relativistic precession starts to dominate the precession rate. For
high eccentricities this happens farther away from the MBH then for
small eccentricities.}
\end{figure}

In Fig. \ref{f:Ta} we show $T_{\rm RR}(a,e)$ as a function of $a$
for several choices of $e$. For large $a$, precession is dominated by
mass precession, and $T_{\rm RR}\propto a$. Closer to the MBH, at a
distance of $\sim0.01\pc$, general relativistic precession starts to
dominate. This happens at larger $a$ when $e$ is large. Near the
minimum, general relativistic precession and precession due to the
extended cluster of stars cancel (Eq. \ref{e:tos}), and resonant
relaxation is limited by the precession rate of the ambient
cluster. For yet smaller $a$, resonant relaxation becomes limited by
general relativistic precession. At distances of $\sim0.01\pc$, the
resonant relaxation time becomes for high eccentricities as small as a
few$\times10^7\yr$.

\section{Summary and discussion}\label{s:disc}

In this paper, we have shown that the net torque of a cluster of stars
on a test star of eccentricity $e$, is proportional to $e$
(Eq. \ref{e:beta_s}). From this dependence, and the dependence of the
precession time on eccentricity, we determine for the first time the
resonant relaxation time as a function of $e$ and $a$ (equations
\ref{e:TRR} and \ref{e:tp}).

Resonant relaxation may play an important role in several phenomena
near MBHs. \citet{Rau96} and \citet{Rau98} estimated that resonant
relaxation may increase the rate of tidal disruptions of stars by the
MBH by a factor $\sim2$ due to the increased rate at which stars are
driven towards the loss-cone. It has also been suggested that resonant
relaxation has modified the distribution of the young star cluster known
as the ``S-stars'' in the Galactic centre \citep{Lev06, Hop06,
Per07}. 

Resonant relaxation also plays a role in the formation of
gravitational wave sources. Compact remnants that spiral into MBHs due
to the emission of gravitational waves are an important potential
source of gravitational waves for the {\it Laser Interferometer Space
Antenna} ({\it LISA}). With the exception of the Galactic centre
(Hopman, Freitag \& Larson 2007), such extreme mass ratio
gravitational wave sources are not observable until they orbit on very
tight orbits with periods less than an hour. Since such stars
originate from orbits relatively close ($\sim 0.01\pc$) to the MBH
\citep{Hop05}, resonant relaxation plays an important role in the
event rate, and can lead to an increase of nearly an order of
magnitude \citep{Hop06}. With the exception of \citet{Fre01, Fre03},
estimates of the event rate have relied on semi-analytical models
which were not fully two dimensional in $(E,J)$-space. In particular,
\citet{Hop06} treated the resonant relaxation time as averaged over
eccentricities.

The determination in this paper of $T_{\rm RR}(a,e)$ as a function of
semi-major axis {\it and} eccentricity, allows for implementation of
resonant relaxation in Monte Carlo codes such as those presented in
\citet{Fre01b, Fre02}. The fact that the torques depend on $e$ can be
of considerable importance for these results. In a companion paper
(Hopman \& G\"urkan 2007) we use the eccentricity dependence derived
in this paper to find the steady state angular momentum distribution
of stars in presence of resonant relaxation, and address the
consequences for the processed mentioned here.

\section*{Acknowledgments}

We thank Tal Alexander, Marc Freitag and Yuri Levin for
helpful discussions, and Ann-Marie Madigan for comments on the
manuscript. M.A.G. was supported by a Marie Curie Intra-European
fellowship under the sixth framework programme, and C.H. by a Veni
scholarship from the Netherlands Organization for Scientific Research
(NWO). C.H. thanks the University of Amsterdam, where most of the work
was done, for their hospitality. The computations in this paper are
done at the Lisa cluster at SARA supercomputing centre in Amsterdam.

\bibliography{/home/clovis/BIBTEX/MasterRefs}

\begin{thebibliography}{}

\bibitem[\protect\citeauthoryear{{Alexander}}{{Alexander}}{1999}]{Ale99a}
{Alexander} T.,  1999, \apj, 527, 835

\bibitem[\protect\citeauthoryear{{Alexander}}{{Alexander}}{2005}]{Ale05}
{Alexander} T.,  2005, \physrep, 419, 65

\bibitem[\protect\citeauthoryear{{Alexander} \& {Hopman}}{{Alexander} \&
  {Hopman}}{2003}]{Ale03b}
{Alexander} T.,  {Hopman} C.,  2003, \apjl, 590, L29

\bibitem[\protect\citeauthoryear{{Bahcall} \& {Wolf}}{{Bahcall} \&
  {Wolf}}{1976}]{Bah76}
{Bahcall} J.~N.,  {Wolf} R.~A.,  1976, \apj, 209, 214

\bibitem[\protect\citeauthoryear{{Bahcall} \& {Wolf}}{{Bahcall} \&
  {Wolf}}{1977}]{Bah77}
{Bahcall} J.~N.,  {Wolf} R.~A.,  1977, \apj, 216, 883

\bibitem[\protect\citeauthoryear{{Chandrasekhar}}{{Chandrasekhar}}{1943}]{Cha4%
3}
{Chandrasekhar} S.,  1943, \apj, 97, 255

\bibitem[\protect\citeauthoryear{{Eisenhauer} et~al.,}{{Eisenhauer}
  et~al.}{2005}]{Eis05}
{Eisenhauer} F.,  et~al., 2005, \apj, 628, 246

\bibitem[\protect\citeauthoryear{{Frank} \& {Rees}}{{Frank} \&
  {Rees}}{1976}]{Fra76}
{Frank} J.,  {Rees} M.~J.,  1976, \mnras, 176, 633

\bibitem[\protect\citeauthoryear{{Freitag}}{{Freitag}}{2001}]{Fre01}
{Freitag} M.,  2001, Classical and Quantum Gravity, 18, 4033

\bibitem[\protect\citeauthoryear{{Freitag}}{{Freitag}}{2003}]{Fre03}
{Freitag} M.,  2003, \apjl, 583, L21

\bibitem[\protect\citeauthoryear{{Freitag}, {Amaro-Seoane} \&
  {Kalogera}}{{Freitag} et~al.}{2006}]{Fre06}
{Freitag} M.,  {Amaro-Seoane} P.,    {Kalogera} V.,  2006, \apj, 649, 91

\bibitem[\protect\citeauthoryear{{Freitag} \& {Benz}}{{Freitag} \&
  {Benz}}{2001}]{Fre01b}
{Freitag} M.,  {Benz} W.,  2001, \aap, 375, 711

\bibitem[\protect\citeauthoryear{{Freitag} \& {Benz}}{{Freitag} \&
  {Benz}}{2002}]{Fre02}
{Freitag} M.,  {Benz} W.,  2002, \aap, 394, 345

\bibitem[\protect\citeauthoryear{{Gebhardt} et~al.,}{{Gebhardt}
  et~al.}{2003}]{Geb03}
{Gebhardt} K.,  et~al., 2003, \apj, 583, 92

\bibitem[\protect\citeauthoryear{{Genzel} et~al.,}{{Genzel}
  et~al.}{2003}]{Gen03a}
{Genzel} R.,  et~al., 2003, \apj, 594, 812

\bibitem[\protect\citeauthoryear{{Ghez}, {Becklin}, {Duchjne}, {Hornstein},
  {Morris}, {Salim} \& {Tanner}}{{Ghez} et~al.}{2003}]{Ghe03b}
{Ghez} A.~M.,  {Becklin} E.,  {Duchjne} G.,  {Hornstein} S.,  {Morris} M.,
  {Salim} S.,    {Tanner} A.,  2003, Astronomische Nachrichten Supplement, 324,
  527

\bibitem[\protect\citeauthoryear{{H\'enon}}{{H\'enon}}{1973}]{Hen73}
	{H\'enon} M.,  1973, in {Contopoulos} G.,  {H\'enon} M.,
	{Lynden-Bell} D.,  eds, ``Dynamical Structure and Evolution
	of Stellar Systems'', 183

\bibitem[\protect\citeauthoryear{{Hopman} \& {Alexander}}{{Hopman} \&
  {Alexander}}{2005}]{Hop05}
{Hopman} C.,  {Alexander} T.,  2005, \apj, 629, 362

\bibitem[\protect\citeauthoryear{{Hopman} \& {Alexander}}{{Hopman} \&
  {Alexander}}{2006a}]{Hop06}
{Hopman} C.,  {Alexander} T.,  2006a, \apj, 645, 1152

\bibitem[\protect\citeauthoryear{{Hopman} \& {Alexander}}{{Hopman} \&
  {Alexander}}{2006b}]{Hop06b}
{Hopman} C.,  {Alexander} T.,  2006b, \apjl, 645, L133

\bibitem[\protect\citeauthoryear{{Ivanov}, {Polnarev} \& {Saha}}{{Ivanov}
  et~al.}{2005}]{Iva05}
{Ivanov} P.~B.,  {Polnarev} A.~G.,    {Saha} P.,  2005, \mnras, 358, 1361

\bibitem[\protect\citeauthoryear{{Levin}}{{Levin}}{2006}]{Lev06}
{Levin} Y.,  2006, ArXiv Astrophysics e-prints

\bibitem[\protect\citeauthoryear{{Lightman} \& {Shapiro}}{{Lightman} \&
  {Shapiro}}{1977}]{Lig77}
{Lightman} A.~P.,  {Shapiro} S.~L.,  1977, \apj, 211, 244

\bibitem[\protect\citeauthoryear{{Miller}}{{Miller}}{2006}]{Mil06}
{Miller} M.~C.,  2006, \mnras, 367, L32

\bibitem[\protect\citeauthoryear{{Murphy}, {Cohn} \& {Durisen}}{{Murphy}
  et~al.}{1991}]{Mur91}
{Murphy} B.~W.,  {Cohn} H.~N.,    {Durisen} R.~H.,  1991, \apj, 370, 60

\bibitem[\protect\citeauthoryear{{Perets}, {Hopman} \& {Alexander}}{{Perets}
  et~al.}{2007}]{Per07}
{Perets} H.~B.,  {Hopman} C.,    {Alexander} T.,  2007, \apj, 656, 709

\bibitem[\protect\citeauthoryear{{Rauch} \& {Ingalls}}{{Rauch} \&
  {Ingalls}}{1998}]{Rau98}
{Rauch} K.~P.,  {Ingalls} B.,  1998, \mnras, 299, 1231

\bibitem[\protect\citeauthoryear{{Rauch} \& {Tremaine}}{{Rauch} \&
  {Tremaine}}{1996}]{Rau96}
{Rauch} K.~P.,  {Tremaine} S.,  1996, New Astronomy, 1, 149

\bibitem[\protect\citeauthoryear{{Rees}}{{Rees}}{1988}]{Ree88}
{Rees} M.~J.,  1988, \nat, 333, 523

\bibitem[\protect\citeauthoryear{{Sch{\"o}del} et~al.,}{{Sch{\"o}del}
  et~al.}{2002}]{Sch02}
{Sch{\"o}del} R.,  et~al., 2002, \nat, 419, 694

\bibitem[\protect\citeauthoryear{{Shapiro} \& {Marchant}}{{Shapiro} \&
  {Marchant}}{1978}]{Sha78}
{Shapiro} S.~L.,  {Marchant} A.~B.,  1978, \apj, 225, 603

\bibitem[\protect\citeauthoryear{{Sigurdsson}}{{Sigurdsson}}{1997}]{Sig97a}
{Sigurdsson} S.,  1997, Classical and Quantum Gravity, 14, 1425

\end{thebibliography}
\bsp
\label{lastpage}
\end{document}